%% file: main.tex
\documentclass[12pt]{llncs}

\usepackage[utf8]{inputenc} 
\usepackage{amsfonts}
\usepackage{graphtex}
\usepackage{epsfig,color}

\input{macros}

\begin{document}

\title{ Peer to Peer Optimistic Collaborative Editing on XML-like Trees}

\author{
Denis Lugiez and Stéphane  Martin 
}

\institute{Laboratoire d'Informatique Fondamentale
 39 rue F. Joliot-Curie,
 13013 Marseille, France
 \{ denis.lugiez, stephane.martin\}@lif.univ-mrs.fr
}

\date{}
\maketitle


\begin{abstract}
Collaborative editing consists in editing a common document shared by several
independent sites. Conflicts occurs when different users perform simultaneous
uncompatible operations. Centralized systems solve this problem by using locks
that prevent some modifications to occur and leave the resolution of conflicts
to users. Optimistic peer to peer (P2P) editing doesn't allow locks and uses a
Integration Transformation IT that reconciliates conflicting operations and
ensures convergence (all copies are identical on each site).  Two properties
TP1 and TP2, relating the set of allowed operations \OP~ and the
transformation IT, have been shown to ensure convergence. The choice of the
set \OP~ is crucial to define an integration operation that satisfies TP1 and
TP2.  Many existing algorithms don't satisfy these properties and are
incorrect. No algorithm enjoying both properties is known for strings and
little work has been done for XML trees in a pure P2P framework.  We focus on
editing XML-like trees, i.e. unranked-unordered labeled trees also considered
in the Harmony project. We show that no transformation satisfying TP1 and TP2
exists for a first set of operations but that TP1 and TP2 hold for a richer
set of operations, provided that some decoration is added to the tree. We show
how to combine our approach with any convergent editing process on strings to
get a convergent process. We have implemented our transformation using a P2P
algorithm inspired by Ressel et al. whose correctness relies on underlying
partial order structure generated by the dependence relation on operations.

\end{abstract}

\section{Introduction}
\label{sec:intro}


Collaborative edition is a concurrent process that allows separate users
-sites- to work on the same data called {\em a collaborative object} using a
set of defined operations. Distinct authors working on the same article,
shared calendar, on-line encyclopedia are example of such processes. This
activity can be centralized by a distinguished site that coordinates and
resolves the conflicts that can arise from concurrent access to the same
resource -for instance two sites want to insert two distinct character at the
same position in a word-, like in the subversion system (svn). A more
liberal approach relies on a peer to peer process (in short P2P) where the set
of users in not fixed in advance and where no central site coordinates the
process. Therefore conflict resolution is much more complex, especially when
one has an optimistic approach that considers that  each operation is
meaningful and must be taken into account. A simpler solution that 
relies on priority attributed to users and undoing conflicting operations can
lead to a situation where only the operations of one user are
performed and  all other operations are discarded,
which is the opposite of a cooperative work. Therefore a main issue in
collaborative edition is to ensure convergence (i.e. each user gets the same
copy of the shared data) in the optimistic framework. The Integration
Transformation approach uses a operator $\IT$ that combines concurrent
operations to get a new operation merging  the effect of these concurrent
operations to resolve the conflicts. Convergence is proved when this
transformation enjoys two properties TP1 and TP2. The problem is
hard for linear structures like words and most algorithms proposed
\cite{ResselNG-CSCW96,SuleimanCF-GROUP97,LiL-ICDCS04} are
non-trivial. Unfortunately recent works \cite{ImineROM-TCS06} show that these
algorithm don't have the convergence property. Furthermore, few results have
been obtained for tree-like structures in a pure P2P optimistic framework
which is the basis for collaborative edition on XML-documents (the solution in
\cite{XMLS06} uses time-stamp, i.e. a central server).  In this paper we
concentrate on labeled unranked-unordered trees, called XML-like trees- which
are already considered in the Harmony project
\cite{HarmonyOverview-TR03} and also provides  a
close approximation to XML-documents (in many applications, the ordering on
siblings on XML document is not relevant). Our first results states that no
$\IT$ transformation can exist for a first basic set of operations. Then we
refine the data structure and we give a rich set of operations that allows to
define an $\IT$ transformation satisfying TP1 and TP2. The proof has been
automated with the Vote system \cite{Imine-PhD06} which uses Spike, a
theorem prover based on term rewriting. Then we show how to combine this data
structure with another date structure for which a convergent algorithm exists
to get convergence for the composed data structure. This results allows
collaborative editing on a complex data structure combining a tree-like
structure and other basic structure like words.

Section \ref{sec:collab-editing} gives the basic definitions, section
\ref{sec:words-and-trees}  describes the main basic data structures words and
trees. Then we give the negative results for these collaborative objects in
section \ref{sec:negative-results}. The new tree-like collaborative object is
given in section \ref{sec:tree-revisited} as well as an integration
transformation  that ensures convergence. Combination of convergent algorithm
are given in Section
\ref{sec:XML-like-trees} and Section \ref{sec:implementation} discusses
implementation issues.


\section{The Framework}
\label{sec:collab-editing}

\subsection{Collaborative Editing and Convergence}
\label{subsec:editing}

A collaborative object consists of a type (calendar, XML document,\ldots) that
defines the set of {\em states}, a set $\Op$ of operations and an operator
$Do$ that applies an operation $op$ to a state $s$ (i.e. an element of the
type) to get another state $op(s)$ that is denoted by $Do(s,op)$. For
instance, the collaborative object {\em word} consists of $\Sigma^*$ the set
of words on an alphabet $\Sigma$, operations {\em InsCh(p,c)} to insert
character $c$ at position $p$, and {\em DelCh(p)} to delete the character at
position $p$ and $Do$ operation simply applies these deletion or insertion to
the current state (which is some word).  A sequence of operations is called an
{\em history} and denoted by $[op_1;op_2;\ldots;op_n]$ and we use the notation
$[op_1;op_2;\ldots;op_n](s)$ to denote
$Do(\ldots,Do(Do(s,op_1),op_2),\ldots,op_n)$ (apply $op_1$ first, then
$op_2$,\ldots).

Collaborative editing is a special kind of concurrent programming on a shared
collaborative object shared by distinct sites.  Centralized systems like svn
have a system of locks that prevent conflicts\footnote{a user can be in
conflict with the master copy, but conflict resolution is under user's
responsibility}, but pure P2P systems have no centralization process that
enforce each site to have the same data. The optimistic approach assumes that
no operation is lost and the main issue is to ensure convergence, i.e. all
sites eventually have the same copy of the shared object.

\paragraph{Requests and computations.}

Each site generate local requests that consists of some operation $op$ to
execute on the shared object plus additional information (site identifier,
operation number, history,\ldots). Each local request is broadcast to all
other sites and we assume that no messages is lost and that the execution
ordering doesn't exchange messages. Requests generated and received by each
site are queued and extracted from the queue to be executed, i.e. the
operation is performed on the current copy of the collaborative object. Local
requests are linearly ordered and the execution of requests respects this
ordering. Therefore requests can be causally related or concurrent (requests
generated independently by distinct sites)

\paragraph{The causality relation and concurrent request.}

Let $r_1^i$ be generated by site $i$ and $r_2^j$ generated by site $j$. The
causality relation $\succ$ is defined by $r_1^i\succ r_2^j$ iff either $i=j$
and $r_1^i$ is generated before $r_2^j$ or $i\neq j$ and the request $r_1^i$
is executed on site $j$ before $r_2^j$. The relation $\succ$ is a partial
order and we say that two requests $r$ and $r'$ are {\em concurrent}, denoted
by $r\parallel r'$, iff $r\not \succ r'$ and $r'\not\succ r$. In the
following, we identity a request and the operation it conveys, and we extend
$\succ$ to operations.


Concurrency may lead to conflicts: For instance two distinct sites insert
different characters at the same position. These conflicts are solved using a
transformational approach. Assume that a site $s$ has performed operation $op$
and that it receives an request containing operation $op'$ that has been
issued by another site $s'$ concurrently to $op$ (i.e. $op\parallel
op'$). Instead of executing $op'$, the site $s$ executes $IT(op',op)$, the
transformation of operation $op'$ according to $op$. Meanwhile site $s'$,
which has executed $op'$ and receives a request to execute $op$ will execute
$IT(op,op')$.  

The {\em convergence property} states that all sites share the same copy of
the collaborative object after they have processed all requests.

\subsection{The Integration Transformation and the Convergence Theorem}
\label{subsec:IT-Convergence}

The Integration function $\IT$ takes two operations $op_2$ issued by $site_2$
and $op_1$ issued by $site_1$ and returns a operation $IT(op_2,op_1)\in
\Op$ that $site_1$ executes. Meanwhile $site_2$ executes $IT(op_1,op_2)$. This
integration function $\IT$ is extended to integrate an operation with a set of
concurrent operations (see
\cite{Imine-PhD06}). 
The classical properties required for ensuring convergence are:
\begin{itemize}
\item $TP1$ property states an {\em  equality on  states}\\ 
$[op_1;IT(op_2,op_1)](t)=[op_2;IT(op_1,op_2)](t)$\\

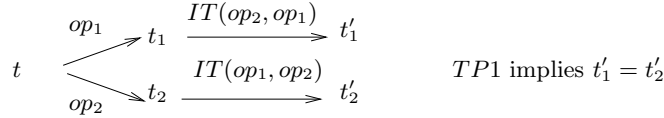
\vspace{-11mm}
\begin{figure}[htbp]
\input{tp1.pstex_t}
\end{figure}

\vspace{-10mm}
\item $TP2$ property states an {\em identity of
operations}:\\ 
$IT(IT(op,op_1),IT(op_2,op_1)) =
IT(IT(op,op_2),IT(op_1,op_2))$
\end{itemize}

\begin{theorem}[\cite{ResselNG-CSCW96}]
If IT satisfies TP1 and TP2 then the convergence property holds.
\end{theorem}

A main issue in collaborative editing is, given a collaborative object, design
an $\IT$ function that satisfies TP1 and TP2. A related issue is to design the
most expressive set of operations, such that there exists an $IT$ satisfying
TP1 and TP2. The larger the set of operations, the better but extending the
set of operations results in a combinatorial explosion when proving TP1 and
TP2. At the present time, no set of operations has been designed to handle
XML-like documents in a pure P2P approach.

\subsection{An Abstract Description of Editing Algorithm}
\label{subsec:abstract-editing-algo}

Each site has a set of local variables $i,s,h,\ldots$ site identifier,
current state of the shared object, history,\ldots and an environment $E$ is a
set of values of these variables (for all sites).  A request is a tuple of
values $\req{i, opnb, op,\ldots}$ (site identifier, operation numbering,
operation,\ldots. The set of environment is $\ENV$ and the set of request is
$\REQ$. 

{\em Local transitions} are described by a transition function $\varphi_l:\Op\times
\ENV\ra \ENV$ that given an operation $op\in \Op$, a current environment $E$
computes the new environment $ E'$ corresponding to the execution of $op$. The
request $r_l$ sent to other sites is the value of some of the local
variables. This process is described as $\varphi_l.!r_l$.

An {\em external request} $r_e$ is followed by a local computation
$\varphi_e:\REQ\times\ENV \ra \ENV$ updating the local variables (using the
$\IT$ function but possibly other functions). This process is described as
$?r_e.\varphi_e$.  A collaborative editing algorithm on a collaborative object
is described by $\ENV,\REQ,\varphi_l,\varphi_e$ (assuming that transformations
like $\IT$ and possibly other functions are already defined).

Each site performs a non deterministic choice between the two processes and
repeats this computation until all messages are processed. A computation is a
sequence of $\varphi_l(op,E).!r_l$ and $?r_e.\varphi_e(E)$ that results from an
interleaving of the computations on each site respecting the causality
relation.

\section{Words and Tree-like Data Structures}
\label{sec:words-and-trees}
\label{sec:unordered-unranked-trees}

In this section, we recall some known facts on words and set up a first
approach for XML-like trees.


The collaborative object {\em word} is given by the set of words on a finite
alphabet $\Sigma$ and the operations $InsCh(p,c)$ that inserts a character
$c\in\Sigma$ at position $p\in \Pos$, $DelCh(p)$ that deletes the character at
position $p\in \Pos$ and $Nop()$ where $\Pos$ is the set of positions
i.e. sequences of integers. Several Transformations $\IT$ have been defined
but none satisfies both $TP1$ and $TP2$ (see section
\ref{sec:negative-results}). Some variants of this object use slightly a more elaborated
data type and operations to keep  track of operations performed at a given
position or for a given character.


The {\em tree} data structure that we define is already used in the Harmony project
\cite{HarmonyOverview-TR03}. 
Let $\Name$ be a set of names, the set $T$ of unordered unranked edge labeled
trees is defined by the grammar:
\[
\begin{array}[t]{rlll}
T::=&\{\} & &\text{ \ // Empty ~tree}\\ 
|& \lmult n_1 (T), ... , n_m(T)\rmult &    ~~n_i\in \Name, n_i \neq n_j $ if $
i\neq j& // Set ~of ~tree \\
\end{array}
\]
The definition ensures that two edges issued from the same node have different
labels: i.e. a given label occurs at most once on siblings.  Trees are
unordered i.e., for any permutation $\sigma$,  we have that
$\{n_1(t_1),\ldots,n_m(t_m)\}=\{n_{\sigma(1)}(t_{\sigma(1)}),\ldots,n_{\sigma(m)}(t_{\sigma(m)})\}$. In
figures, we draw $\mult{}$ as a node, and we add a root node to a tree
$\mult{n_1(t_1),\ldots,n_m(t_m)}$.

\begin{example}
\[
t= \\ 
\left \{ 
\begin{array}{l} 
Pat \left ( \left\{ \begin{array}{l} 
                     Phone \left ( \left \{ \begin{array}{l} 
                                             Home(\{0491543545(\{\})\}) \\
                                             Cellular(\{0691543545(\{\})\}) 
                                             \end{array}
                                    \right \}  
                           \right )
                     \end{array} 
            \right \} 
      \right )\\ 
Henri(\{ Address(\{45~ Emile~ Caplant~ Street ( \{\})\}) \}) 
\end{array}\right\}
\]

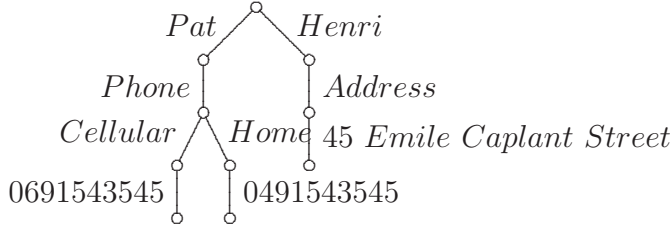
\pic
\Edge  (0,0) (-20,-20) (0,0) (20,-20) (-20,-20) (-20,-40) (20,-20) (20,-40) (-20,-40) (-30,-60) (-20,-40) (-10,-60) (20,-40) (20,-60) (-30,-60) (-30,-80) (-10,-60) (-10,-80)
\Vertex(0,0) (-20,-20) (20,-20) (20,-40) (-20,-40) (-30,-60) (-10,-60) (20,-60) (-30,-80) (-10,-80)

\Align[r] ($Pat$) (-15,-7)
\Align[l] ($Henri $) (15,-7)
\Align[r] ($Phone$) (-25,-30)
\Align[l] ($Home$) (-11,-47)
\Align[r] ($Cellular$) (-30,-47)
\Align[l] ($0491543545 $) (-5,-70)
\Align[r] ($0691543545 $) (-35,-70)
\Align[l] ($Address$) (25,-30)
\Align[l] ($45~ Emile ~Caplant~ Street$) (25,-50)

\cip

\end{example}

A path is a sequence of names, $\epsilon$ is the empty path and $p.p'$ is
the concatenation of paths $p$ and $p'$. The set of paths is written $\mathcal{P}$. The 
projection of tree $t$ along a path $p$, written $\proj{p}{t}$, is defined by
$\proj{\epsilon}{t}=t$ and $\proj{n.p}{t}=\proj{p}{\proj{n}{t}}$, $n\in \Sigma,
p \in \mathcal{P}$.
We write $p_1 \triangleleft p_2$, when a path $p_1$ is a prefix of another
path $p_2$.

The operations that we consider are:


\begin{itemize}
\item $Add(p,n)$ : Add a edge labeled $n$ at end of path $p$.
$$\begin{array}{l}
Add(n'.p,n)(\{n_1(t_1),...,n_q(t_q)\}) =\\
  \hspace{2cm}      \lmult n_1(t_1),...,n_q(t_q),n'(Add(p,n)(\{\}))\rmult 
         $ if $ n'\not \in Dom(t)\\
Add(n_i.p,n)(\{n_1(t_1),...,n_i(t_i),...,n_q(t_q)\})=\\
\hspace{3cm}\lmult n_1(t_1),...,n_i(Add(p,n)(t_i)),...,n_q(t_q)\rmult \\
Add(\epsilon,n)(t)= t , $ if $ n\in Dom(t)\\
Add(\epsilon,n)(\lmult n_1(t_1),...,n_q(t_q)\rmult)
 =\lmult n_1(t_1),...,n_q(t_q),n(\{\})\rmult
\\
\end{array}$$

Example : $t'=Add(Henri.Phone, 0491835469)(t)$

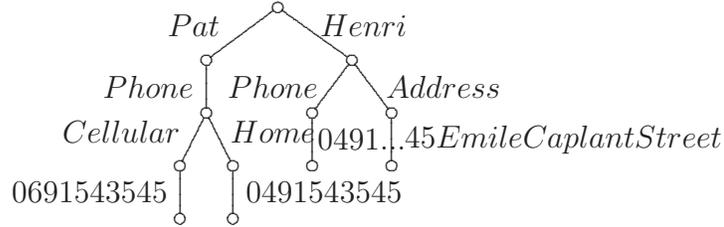
$$ \pic
\Edge  (7,0) (-20,-20) (7,0) (35,-20) (-20,-20) (-20,-40) (35,-20) (20,-40) (-20,-40) (-30,-60) (-20,-40) (-10,-60) (20,-40) (20,-60) (-30,-60) (-30,-80) (-10,-60) (-10,-80) (35,-20) (50,-40) (50,-40) (50,-60)
\Vertex(7,0) (-20,-20) (35,-20) (20,-40) (20,-60) (50,-40) (50,-60) (-20,-40) (-30,-60) (-10,-60)  (-30,-80) (-10,-80) 
%
\Align[r] ($Pat$) (-15,-7)
\Align[l] ($Henri $) (23,-7)
\Align[r] ($Phone$) (-25,-30)
\Align[l] ($Home$) (-11,-47)
\Align[r] ($Cellular$) (-30,-47)
\Align[l] ($0491543545 $) (-5,-70)
\Align[r] ($0691543545 $) (-35,-70)
\Align[l] ($Address$) (48,-30)
\Align[l] ($45 Emile Caplant Street$) (55,-50)
\Align[r] ($Phone$) (22,-30)
\Align[l] ($0491...$) (22,-50)
\cip
$$
 $Add(Henri,Phone)(t')=t'$ since $Henri.Phone$ already exists.
\item $Nop()$ : Do nothing. $Nop()(t)=t$
\item
$Del_1(p,n)$: Replace a edge labeled $n$ at end of path $p$ by the set of its
successors.\\ 
$$
\begin{array}{l}
Del_1(n'.p,n)(t)=t, $ if $ n\not \in Dom(t)\\
Del_1(n_i.p,n)(\lmult n_1(t_1),...,n_i(t_i),...,n_q(t_q)\rmult) =\\
   ~~~~~\lmult n_1(t_1),...,n_i(Del_1(p,n)(t_i)),...,n_q(t_q)\rmult\\
Del_1(\epsilon,n)(t)=t , $ if $ n\not \in Dom(t)\\
Del_1(\epsilon,n_i)(\{n_1(t_1),...,n_i(t_i),...,n_q(t_q)\})=
                \lmult n_1(t_1),...,n_q(t_q)\rmult \oplus t_i
\end{array}
$$

\end{itemize}

\section{Negative Results for Words and Trees}
\label{sec:negative-results}

\paragraph{The Word Case.}

Imine's work \cite{Imine-PhD06,ImineROM-TCS06} contains counter-examples for the
convergence property for the algorithms presented in
\cite{ResselNG-CSCW96,SuleimanCF-GROUP97,LiL-ICDCS04} and discusses this
issue. He  defines a weaker property TP2' which requires the identity on states
instead of operations. Then he gives an algorithm ensuring convergence relying
on TP2' but this algorithm 
needs the reordering of histories. Therefore, we can state:

\begin{proposition}[\cite{Imine-PhD06}]
No  transformation $\IT$ for words described in the literature satisfies TP1 and TP2.
\end{proposition}

\paragraph{Unordered Unranked Trees.}

Let $\Op=\{Nop(),Add(p,n),Del_1(p,n)\}$ $p\in
\mathcal{P}, n \in \Sigma$. 
We say that an operation $op(x_1,\ldots,x_n)$ is defined from $\Op$ iff
$op(x_1,\ldots,x_n)$ is some element of $ Exp$ according to the grammar:
\[
\begin{array}{lcll}
Exp&::=&op(y_1,\ldots,y_p)~|~$ if $Cond$ then $Exp$ else $Exp2$ fi$& op\in \Op\\
Cond&::= &x \bowtie y~|~Cond \et Cond~|~\non Cond&\\
\end{array}\\
\]
where $\bowtie$ denotes $=$ or $\lhd$, $x,y$ are variables or expressions $p.n$.
This grammar capture the natural definitions of any operation on trees from
the basic operations of $\Op$ excepting iteration and recursion which are
out of scope  in our  framework.

\begin{theorem}
\label{thm:no-IT-TP1-XML}
There is no definition of $IT(op_1,op_2)$ from $Op$ such that $IT$ satisfies $TP1$. 
\end{theorem}



We can restore $TP1$ and $TP2$ using a stronger notion of deletion (See Appendix \ref{subsec:delete2-is-TP1-TP2}). Let
$Del_2$ be the operation  deleting the entire subtree and let $\Op'=\{Nop(),Add(p,n),Del_2(p,n)\}$ $p\in
\mathcal{P}, n \in \Sigma$.

\begin{theorem}
\label{thm:trees-IT-TP1-TP2}
There is a $IT$ for $\Op'$ that  satisfies $TP1$ and $TP2$.
\end{theorem}


\section{Unordered Unranked Trees Revisited}
\label{sec:tree-revisited}

In collaborative edition each site is identified by its number and numbers the
operations that it performs. This ordering is linear and unambiguous.  When a
tree is constructed from the empty tree, one can uniquely label each edge by
the site number and the numbering of the operation that has created this
edge. Since we can also add labels like those of XML-documents, we have a data
structure that corresponds to unordered XML documents where the edges are
labeled by an item occurring once in the tree.


\subsection{The Data Structure}

A identifier is either one of the reserved names {\em doc} (for document) or
{\em mem} (for memory) or a pair of natural numbers $(site,nbop)$ where the
$site$ denotes a site number and $nbop$ denotes an operation number.  $\ID$
denotes the set of identifiers.  A label $l$ is an element of a set of labels
$\L$ (for instance {\em section, paragraph,\ldots}).

We consider trees defined as in section \ref{sec:unordered-unranked-trees} on the set
of names $\Name=\L\times \ID$ assuming that each identifier occurs once in the tree.
\[
T::=\{\} ~|~ \{(l_1,id_1) (T), ... , (l_m,id_m)(T)\}
\]
where each $id_i$'s occurs once in the whole tree.
\begin{example}
\begin{small}
\[
t= 
\left \lmult \begin{array}{l} 
(Pat,1;1) 
 \left ( \left \lmult \begin{array}{l} 
                      (Phone,2;1) \left ( \left \lmult \begin{array}{l} 
                                                        (Home,3;1)(\lmult(0491543545,4;1)(\lmult\rmult)\rmult) \\
                                                        (Cellular,5;1)(\lmult(0691543545,6;1)(\lmult\rmult)\rmult)                               \end{array}
                                           \right \rmult  
                                   \right )
                       \end{array} 
          \right \rmult 
   \right )\\
(Henri,2;2)(\lmult (Address,3;2)(\lmult (45~ Emile~ Caplant~ Street,4;2)
(\lmult\rmult)\rmult) \rmult) 
            \end{array}
\right \rmult
\]
\end{small}
\end{example}
From now on, $\munion$ denotes the union of multisets. Actually, we use this
operation only for disjoint sets, computing a set (not a multiset).  We define
two projection operations:

The projection $\proj{id}{t}$  of $t$ in $id$ is defined by
$\proj{id_i}{\lmult\rmult} = \lmult\rmult$ and\\
\begin{array}[t]{l}
\proj{id_i}{\lmult(l_1,id_1)(t_1),...,(l_i,id_i)(t_i),...,(l_m,id_m)(t_m)\rmult}
=t_i\\
\proj{id}{\lmult(l_1,id_1)(t_1),...,(l_m,id_m)(t_m)\rmult}
=\proj{id}{t_1}\munion...\munion\proj{id}{t_m}\\
\end{array}\\
and  the second projection $\projb{id_i}{t}$ is defined by
$\projb{id_i}{\lmult\rmult} = \lmult\rmult$ and \\
\begin{array}[t]{l} 
\projb{id_i}{\lmult(l_1,id_1)(t_1),...,(l_i,id_i)(t_i),...,(l_m,id_m)(t_m)\rmult}
=\lmult(l_i,id_i)(t_i)\rmult\\
\projb{id}{\lmult(l_1,id_1)(t_1),...,(l_m,id_m)(t_m)\rmult}
=\lmult \projb{id}{t_1},...,\projb{id}{t_m}\rmult\\
\end{array}

\begin{example}
Let $t$ be as above, then:
\[
\begin{array}[t]{l}
\proj{3;1}{t}  =\lmult(0491543545,4;1)(\lmult\rmult)\rmult \\
\projb{3;1}{t} =\lmult(Home,3;1)(\lmult(0491543545,4;1)(\lmult\rmult)\rmult)\rmult
\end{array}
\]
\end{example}

Each tree can be transformed into an (unordered) XML tree by the tree morphism
defined by $\T2XML(\mult{})=\mult{}$ and\\
\begin{array}[t]{l}
\T2XML(\lmult(l_1,id_1)(t_1),\ldots,(l_m,id_m)(t_m) \rmult)= 
\lmult (l_1)(\T2XML(t_1)),\ldots,(l_m)(\T2XML(t_m)) \rmult
\end{array}

\subsection{Gluing Memory and Tree in a Single Tree}

As already mentioned, the collaborative object that we use consists in two
parts: one is a tree that represents the document that we edit and the other
one is a memory where we keep some previous parts of the document that have
been erased. The memory is needed because solving conflicts may require to
fetch parts of the trees in the memory to update the document part (this comes
from the move operation $Mv$).  To get a uniform definition for operations, we
represent the memory and the document in a single tree, so-called well-formed
tree. A {\em well-formed tree} is a tree of the form
$\mult{(\bot,\data)(t_d),(\bot,\mem)(t_m)}$ where $\bot$ is some new label.

\paragraph{The Set of Operations $\Op$.}
\label{subsubsec:operations}

Firstly, we define two auxiliary functions: 
\begin{itemize}
\item  $Erase(id,t)$ deletes the node having identifier $id$ in $t$.
$$\begin{array}{l}
Erase(id,\lmult\rmult)=\lmult\rmult\\

Erase(id,\lmult(l_1,id_1)(t_1),...,(l_q,id_q)(t_q)\rmult)=\\
~~~~\lmult(l_1,id_1)(Erase(id,t_1)),...,(l_q,id_q)(Erase(id,t_q))\rmult\\ 

Erase(id,\lmult(l_1,id_1)(t_1),...,(l_{id},id)(t_{id}),...,(l_q,id_q)(t_q)\rmult)=\\
~~~~~\lmult(l_1,id_1)(t_1),\ldots,
       (l_{i-1},id_{i-1})(t_{i-1},(l_{i+1},id_{i+1})(t_{i+1}),\ldots,
       (l_q,id_q)(t_q)\rmult

\end{array}$$
\item  $AddTree(id,s,t)$ adds $s$ under identifier $id$ in $t$ (performing
union of $s$ and of the subterm in $t$).
$$\begin{array}{l}
AddTree(\idp,t',\lmult\rmult)=\lmult\rmult\\

AddTree(\idp,t',\lmult(l_1,id_1)(t_1),...,(l_q,id_q)(t_q)\rmult)=\\
~~~~\lmult(l_1,id_1)( AddTree(\idp,t',t_1)),...,(l_q,id_q)( AddTree(\idp,t',t_q))\rmult\\ 

AddTree(id,t',\lmult(l_1,id_1)(t_1),...,(l_{id},id)(t_id),...,(l_q,id_q)(t_q)\rmult)=\\
~~~~\lmult(l_1,id_1)(t_1),...,(l_{id},id)(t_{id}\munion t'),...,(l_q,id_q)(t_q)\rmult
\end{array}$$
\end{itemize}

Let 
$\Op=\{Add(\idp,n,id),Del(id),Mv(id,\idp),Ren(id,n),Nop()\}$, 
where $id\in \ID \setminus
\{\data,\mem\},\idp\in \ID, n\in \Sigma$ be the new set of operations.

\begin{itemize}

\item $Add(\idp,id)$: Add a edge labeled $noValue$ with identifier  $id$ under
a node whose identifier is $\idp$.
\[
Add(\idp,id)(t)=AddTree(\idp,\lmult(id,NoValue)(\lmult\rmult)\rmult,t )
\]

\item $Del(id)$: Delete a node $id$ and store deleted subtree in memory.
\[
Del(id)(t)= AddTree(\mem,\proj{id}{t},Erase(id,t))
\]

\item $Mv(id,\idp)$: Move node $id$ under node $\idp$ 
$$Mv(id,\idp)(t)=AddTree(\idp, \projb{id}{t},Erase(id,t))$$

\item $Ren(id,n)$: Change label of node $id$
\[
\begin{array}{l}
Ren(id,l)(\lmult\rmult)
=\lmult\rmult\\
Ren(id,l)(\lmult(l_1,id_1)(t_1),...,(l_q,id_q)(t_q)\rmult)=\\
~~~~~~~~~\lmult(l_1,id_1)( Ren(id,l)(t_1)),...,(l_q,id_q)( Ren(id,l)(t_q))\rmult\\ 
Ren(id,l)(\lmult(l_1,id_1)(t_1),...,
(n',id)(t_i),...,(l_q,id_q)(t_q)\rmult)=\\ 
~~~~~~~~~\lmult(l_1,id_1)(t_1),...,(n,id)(t_i),...(l_q,id_q)(t_q)\rmult\\
\end{array}
\]

\item $Nop()$: Do nothing.
$Nop()(t)=t$
\end{itemize}
Besides basic operations for adding and deleting edges, we add two
useful  operations, one for renaming labels (change a \verb#\section# to a
\verb#\subsection# for instance) and another one for moving parts of a tree
(let's move the {\em \verb#\#theorem} before the {\em \verb#\#corollary} for instance). This
last operation is the reason why we need a memory part in the tree.

\begin{proposition}
Let $t$ be a well-formed tree, let $op\in \Op$, then $op(t)$ is a well-formed
tree. 
\end{proposition}

\begin{remark}
By definition an identifier $id$ is created once since it is equal to
$(site,nbop)$ where site is the number of the site which has created it and
$nbop$ is the numbering of the creation operation. Therefore if the edge
corresponding to this identifier is created, and deleted later on, it cannot
be re-created (since the numbering or the site number is different). An edge can
be created at the ``same'' place\footnote{we use the intuitive notion of {\em
same} here}, but with a different identifier.
\end{remark}

\subsection{The $\IT$ Transformation}


\begin{figure}[htbp]
\[
\begin{array}{l}
IT(Add(\idp,id),Add(\idp',id'))= Add(\idp,id), \\
IT(Add(\idp,id),Del(id'))=
   \left \{ 
   \begin{array}{l} 
   Nop() ~if~ id=id'\\
   Add(\mem,id) ~if~  \idp=id'\\
   Add(\idp,id)  ~ otherwise \\
   \end{array} 
   \right.\\ 
IT(Del(id),Add(\idp',id'))=Del(id)\\
IT(Del(id),Del(id'))=Del(id)\\
IT(Ren(id_1,l_1),Ren(id_2,l_2))=
   \left \{ 
   \begin{array}{l} 
   Nop() ~if~  s_2<s_1 \wedge id_1=id_2\\
   Ren(id_1,l_1) ~otherwise.\\
   \end{array} 
   \right.\\ 
IT(Ren(id_1,l_1),op)=Ren(id_1,l_1) ~if~  op \neq Ren \\
IT(op,Ren(id_1,l_1))=op  ~if~   op \neq Ren \\
IT(Mv(id_1,\idp),Mv(id_2,\idp'))=
   \left \{ \begin{array}{l}
             Nop(),$ ~if~ $ s_2<s_1 \wedge id_1=id_2 \\
             Mv(id_1,\idp) ~otherwise\\
             \end{array} 
    \right.\\ 
IT(Mv(id_1,\idp),Del(id'))=
    \left \{ 
    \begin{array}{l}
     Mv(id_1,\mem) ~if~ \idp=id' \\
     Nop() ~if~ id_1=id'\\ 
     Mv(id_1,\idp) ~otherwise\\
     \end{array} 
     \right.\\ 
IT(Mv(id_1,id_2),op)=Mv(id_1,id_2) ~if~  op\neq Mv, Del\\
IT(op,Mv(id_1,id_2))=op  ~if~ op\neq Mv\\
IT(op_1,Nop())=op_1\\
IT(Nop(),op_2)=Nop();\\
\end{array} 
\]
where $\idp,\idp'\in ID, id\in \ID\setminus{\{\data,\mem\}}$.
\caption{The Transformation IT}
\label{fig:it}
\end{figure}

\begin{theorem}
\label{thm:IT-is-TP1-TP2}
The transformation $\IT$ defined in figure \ref{fig:it} satisfies TP1 and TP2.
\end{theorem}

\begin{proof}
The proof relies on a highly combinatorial case analysis and was double
checked using the Vote tool \cite{Imine-PhD06}. 
\end{proof}

\section{Combining XML-like Trees and Words}
\label{sec:XML-like-trees}

\paragraph{Composition of Trees and Words.}

Let $(T,Op_T,Do_T)$ be the collaborative object obtained from trees and the
set of operations defined in section \ref{sec:tree-revisited}. Let $Dom(t)$
be the set of identifier occurring in $t\in T$. Let
$Data=(D,Op_D,Do_D,)$ be another collaborative object. We assume that $d_0\in
D$ is the default initial value for elements of type  $D$.
Let $\delta:ID \ra D$ be a labelling function that associates to each $id\in
\ID$ some element $d=\delta(id)$ of $Data$. A labeled tree is a pair
$t,\delta$ and $T(D)$ denotes the set of labeled trees. For instance the
labelling can associate to each identifier $id$ a string that can be the
information stored at the terminal node of the edge labeled by $id$, we call
this data structure XML-like trees.
 
We define the collaborative object $T(Data)$, the trees parameterized by $Data$,
as follows:

\begin{itemize}
\item The set of states is $T(D)$,

\item The set $Op$ of operations is composed of 
$op_{id}$ for $id\in \ID$, $op_{id}\in Op_D$, 
and $op$ where $op\in Op_T$.



\item The $Do$ function is defined by

$Do((t,\delta),op_{id})=(t,\delta')$ where  the labelling  $\delta$ is
identical to $\delta$ except  that $\delta'(id)=op(\delta(id))$.

$Do((t,\delta),op)=(t',\delta')$ where $t'=Do(t,op)$ and $\delta'$ is
identical to $\delta$ except that $\delta(id)=d_0$ (the default value of $D$)
if $id$ is an identifier not occurring in $t$.

\end{itemize}


\paragraph{Composition of Convergent Algorithms.}

Let $\A_T$ be a convergent collaborative editing algorithm for  $T$ defined
by $\ENVT,\REQT, ?r^T_e.\varphi^T_e, \varphi^T_l.!r^T_l$ and let 
let $\A_D$ be a convergent collaborative editing algorithm for Data defined
by $\ENVD$,$\REQD$, $?r^D_e.\varphi^D_e$, $\varphi^D_l.!r^D_l$.
We define a collaborative editing algorithm for $T(D)$ by composing both
algorithm in a product-like way.
Environments have the form $\pair{E_T,E_D}$ where $E_T\in\ENVT$ and $E_D$ is a
partial function $\ID\ra \ENVD$. The function is defined for $id \in Dom(s)$
where $s\in E_T$ is the state of the collaborative object.  Similarly requests
have the form $\pair{r_T,\bot}$ or $\pair{\bot,r_D}$ where $\bot$ stands for
undefined, $r_T\in \REQT$ and $r_D$ is a pair $(id,r)$ with $id\in \ID, r\in
\REQD$. The set of environment is denoted by $\ENV$, the set of requests is
denoted by $\REQ$.
The composition  is defined by
\begin{itemize}
\item
Local computation $\phi_l:\Op,\ENV\ra \ENV$ where\\
$
\begin{array}{l}
\phi_l(op,\pair{E_T,E_D})= \varphi^T_l(op,E_T)$ and $r_l=\pair{r^T_l,\bot}$ if $op\in Op_T\\
\phi_l(op,\pair{E_T,E_D})= \varphi^D_l(op_{id},E_D(id))$ and $r_l=\pair{\bot,(id,r^D_l)}\\
$ if $op=op_{id}\in Op_D
\end{array}
$
\item
Computation following external requests $\phi_e:\REQ,\ENV\ra \ENV$ where
$\phi_e(r_e,\pair{E_T,E_D})=\varphi_e^T(r_e^T,E_T)$ if $r_e=\pair{r_e^T,\bot}$
and
\\ 
\hspace{-6mm}
$
\begin{array}{l}
\phi_e(r_e,\pair{E_T,E_D})=\varphi_e^D(r_e^D,E_D(id))$ if $E=\pair{E_T,E_D}$, $r_e=\pair{\bot,(id,r_e^D)}\\
\end{array}
$
\end{itemize}
The initial state is the empty tree, labeled by $d_0$ and the current state
is the tree which is the current state $s_T$ computed by $\A_T$ and for each
$id\in Dom(s_T)$ the labelling is the state computed by $A_D$.

\begin{theorem}
\label{thm:combination}
If $A_D$ and $A_T$ are convergent, then their composition is convergent.
\end{theorem}



Let XML-like documents be labeled unranked-unordered trees decorated with
strings. Since convergent algorithms for words exist (more complex than
algorithms using $\IT$, see
\cite{LiL-CollaborateCom05,Imine-PhD06} for instance)  and since the
transformation $\IT$ of section \ref{sec:tree-revisited} is TP1 and
TP2, we have:

\begin{theorem}
There exists a convergent editing algorithm for  XML-like documents.
\end{theorem}

\section{Algorithm and Implementation}
\label{sec:implementation}

The algorithm follows the lines given at section
\ref{subsec:abstract-editing-algo}. It is  similar to
\cite{ResselNG-CSCW96,LushmanC-IPL03}, but we replace the  explicit vector
dependency by sending the set of (minimal dependencies) of the operation sent
by the site.  This amounts to giving an slightly modified version of the
translate function that computes the integration of an operation with respect
to a set of dependencies. Therefore the set of sites is not fixed in advance
and can evolve during the editing process. As mentioned in
\cite{LushmanC-IPL03}, the correctness of this algorithm relies on the partial
ordering structure underlying the set of requests.

The implementation has been done in Java and performs well in
practice. Examining random execution of the algorithm shows that most of the
computations are implicitly independent: operations on nodes of distinct
identifiers don't interfere. The operations that may cause actual conflicts
are renaming of labels (on the same identifier). In many other cases, the
integration $\IT(op,op')$ returns $op$.

We plan to investigate further the algorithm and its properties to give
theoretical bases for a set of optimizations that can improve its efficiency.
For instance, we have proved that integrating an operation with pairwise
disjoint operations always return the same operation, therefore some
memoization techniques could be used to save computation time.






\section{Conclusion}
\label{sec:conc}

We have proposed a first approach to deal with XML-like trees in a P2P
Collaborative Editing framework using a rich set of operations and a
transformation enjoying the key properties to ensure convergence (when none of
existing algorithms for words achieve this goal). We are currently
investigating several issues. The first one is to deal with ordered unranked
trees but, since this case contains the word case, the problem is hard and the
existence of a simple integration transformation is still pending. Another
issue is to deal with typing issues, where the relevant notion of type is
regular tree languages for unordered-unranked tree languages (that generalizes
DTD and XML-Schemas to this data-structure) like in
\cite{FosterPS-PlanX07}. The first results in this direction shows that
requiring to use transformations that respect types strongly restrict the
class of well-typed trees. Finally, trees have a structure which is inherently
concurrent (branches are independent up to their common root) and can be
exploited to improve the computational aspects of our algorithm.



\bibliographystyle{plain}
\bibliography{/home/lugiez/Recherche/MABIBLIO/abbrev,/home/lugiez/Recherche/MABIBLIO/revues,/home/lugiez/Recherche/MABIBLIO/conferences,/home/lugiez/Recherche/MABIBLIO/rapports,/home/lugiez/Recherche/MABIBLIO/theses,/home/lugiez/Recherche/MABIBLIO/livres,/home/lugiez/Recherche/MABIBLIO/misc}

\pagebreak

\section*{Appendix}

\subsection{Proof of Theorem \ref{thm:no-IT-TP1-XML}}

<<<<<<< main.tex

\begin{proof}
The proof is by induction on $n$.

\begin{itemize}
\item Base case $n=1$. The result is obvious (the only substitution is the identity.

\item Inductive step. We assume that for all $op, op_1,\ldots,op_{n-1}$,
$\sigma$ permutation of $\{1,\ldots,n-1\}$ we have
$\ITS(op,[op_1,\ldots,op_{n-1}])=\ITS(op,[op_{\sigma(1)},\ldots,op_{\sigma(n-1)}])$.

Let $op, op_1,\ldots,op_{n}\in \Op$ and let $\sigma$ be a permutation of
$\{1,\ldots,n\}$. We distinguish several cases:

\begin{itemize}
\item $\sigma(n)=n$. Then $\sigma$ is a permutation of $\{1,\ldots,n-1\}$.

\[
\begin{array}[t]{lcl}
\ITS(op,[op_{\sigma(1)},\ldots,op_{\sigma(n)}])&=&
               \ITS(op,[op_{\sigma(1)},\ldots,op_{\sigma(n-1)},op_n])\\
&=&\IT(
      \begin{array}[t]{l}
      \ITS(op,[op_{\sigma(1)},\ldots,op_{\sigma(n-1)}]),\\
      \ITS(op_n,[op_{\sigma(1)},\ldots,op_{\sigma(n-1)}])
      \end{array}\\
&=&\IT(\begin{array}[t]{l}
       \ITS(op,[op_{\sigma1},\ldots,op_{n-1}]),\\
       \ITS(op_n,[op_{1},\ldots,op_{n-1}]))
       \end{array}\\
&&(by ~induction~ hypothesis)\\
&=&\ITS(op,[op_1,\ldots,op_{n-1},op_n])\\
\end{array}
\]

\item $\sigma$ exchanges $n$ and $n-1$ and $\sigma(i)=i$ for $i\neq n,n-1$.

\[
\begin{array}[t]{lcl}
\ITS(op,[op_{\sigma(1)},\ldots,op_{\sigma(n)}])&=&
               \ITS(op,[op_{1},\ldots,op_{n-2},op_n,op_{n-1}])\\
&=&\IT(
       \begin{array}[t]{l}
       \ITS(op,[op_{1},\ldots,op_{n-2},op_{n}]),\\
       \ITS(op_{n-1},[op_{1},\ldots,op_{n-2},op_{n}]))
       \end{array}\\
&=&\IT( 
      \begin{array}[t]{ll}
      \IT&(\ITS(op,[op_{1},\ldots,op_{n-2}]),\\
          &\ITS(op_n,[op_1,\ldots,op_{n-2}])),\\
      \IT&(\ITS(op_{n-1},[op_{1},\ldots,op_{n-2}]),\\
         & \ITS(op_{n},[op_{1},\ldots,op_{n-2}])))
       \end{array}\\
&=&\IT(\IT(op,op_1),\IT(op_2,op_1)\\
&=&\IT(\IT(op,op_2),\IT(op_1,op_2)~(by ~TP2)\\
&=&\IT(\begin{array}[t]{l}
       \IT(\ITS(op,[op_{1},\ldots,op_{n-2}]),\\
        ~~~~~\ITS(op_{n-1},[op_{1},\ldots,op_{n-2}])),\\
       \IT(\ITS(op_n,[op_1,\ldots,op_{n-2}]),\\
        ~~~~~\ITS(op_{n-1},[op_{1},\ldots,op_{n-2}]))))
       \end{array}\\
&=&\ITS(op,[op_1,\ldots,op_{n-2},op_{n-1},op_n])\\
\end{array}
\]

\item $\sigma(n)\neq n,n-1$. Then $\sigma$ can be composed as three
substitutions $\sigma_1,\sigma_2,\sigma_3$:

$\sigma_1$ exchanges $n-1$ and $\sigma(n)$  and leave
other element unchanged (hence $\sigma_1(n)=n$ since $\sigma(n)\neq n$).
$\sigma_2$ exchanges $n-1$ and $n$.
$\sigma_3(n)=n$ and $\sigma_3$ is such that
$\sigma(i)=\sigma_3(\sigma_2(\sigma_1(i)))$. 

By the first case 
\[
\ITS(op,[op_1,\ldots,op_n])=\ITS(op,[op_{\sigma_1(1)},\ldots,op_{\sigma_1(n)}])\]
By the second case
\[
\ITS(op,[op_{\sigma_1(1)},\ldots,op_{\sigma_1(n)}])=
\ITS(op,[op_{\sigma_2(\sigma_1(1))},\ldots,op_{\sigma_2(\sigma_1(n))}])\]
By the first case again 
\[
\ITS(op,[op_{\sigma_2(\sigma_2(\sigma_1(1))},\ldots,op_{\sigma_2(\sigma_1(n))}])
=\ITS(op,[op_{\sigma_3(\sigma_2(\sigma_2(\sigma_1(1)))},\ldots,op_{\sigma_3(\sigma_2(\sigma_1(n)))}])
\]
Therefore
\[
\ITS(op,[op_1,\ldots,op_n])=\ITS(op,[op_{\sigma(1)},\ldots,op_{\sigma(n)}])
\]
\end{itemize}

\end{itemize}

\end{proof}













\begin{proposition}

\end{proposition}

\begin{proof}
\end{proof}

\subsection{Proof of Theorem \ref{thm:no-IT-TP1-XML}}
=======
We prove that no $\IT$ exists for our first set of operations on trees.
>>>>>>> 1.3

\begin{proof}
We assume that $TP1$ holds and we prove that $IT(op_1,op_2)$ can't be defined
on an operation of $Op$.  Let $t_1=op_1(t)$, $t_2=op_2(t), t'_1=op'_2(t_1)$
with $op'_2=IT(op_2,op_1)$, $t'_2=op_1'(t_2)$ with $op'_1=IT(op_1,op_2)$. We
assume that $IT(op_1,op_2)$ is another operation of \OP. The extension to a
boolean combination of operation is straightforward.

\begin{itemize}
\item $op_2'=Nop()$
\begin{itemize}
\item $op_1'=Nop()$ : Trivial because $t_1 \neq t_2$
\item $op_1'=Add(\_,\_)$ \\
Then there is at least one more edge on $t'_2$.
\item $op_1'=Del(x,y)$ we get : 
$$
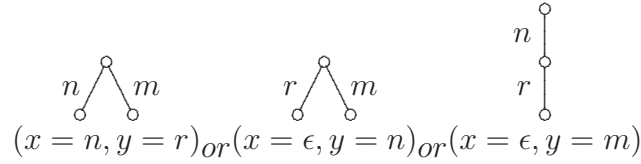
\pic 
\Vertex(0,0) (-10,-20) (10,-20) 
\Edge (0,0) (-10,-20) (0,0) (10,-20)  
\Align[r] ($n$) (-10,-10)
\Align[l] ($m$) (10,-10)
\Align[c] ($(x=n,y=r)$) (0,-30)
\cip
 or 
\pic 
\Vertex(0,0) (-10,-20) (10,-20) 
\Edge (0,0) (-10,-20) (0,0) (10,-20) 
\Align[r] ($r$) (-10,-10)
\Align[l] ($m$) (10,-10)
\Align[c] ($(x=\epsilon,y=n)$) (0,-30)
\cip
or
\pic 
\Vertex(0,0) (0,-20) (0,-40)
\Edge (0,0) (0,-20)  (0,-20) (0,-40) 
\Align[r] ($n$) (-5,-10)
\Align[r] ($r$) (-5,-30)
\Align[c] ($(x=\epsilon,y=m)$) (0,-50)
\cip$$
Any possible operation leaves $t_2$ unchanged.\\

In all case $t_1'\neq t_2'$
\end{itemize}
\item $op_2'=Add(\_,\_)$
\begin{itemize}
\item $op_1'=Nop()$\\
We have $r$ under $m$ on $t_1$ and under $n$ on $t'_2$.
\item $op_1'=Add(\_,\_)$ \\
The number of edges on $t'_1$ and on $t'_2$ are different.
\item $op_1'=Del(\_,\_)$ 
same case
\end{itemize}
\item $op_2'=Del(\_,\_)$
$$\pic
\Vertex(0,0) (0,-20) (30,0) (30,-20)
\Edge (0,0) (0,-20) (30,0) (30,-20)
\Align[r] ($t_1'=m$) (-5,-10)
\Align[r] ($r$) (25,-10)
\Align[c] (or) (10,-10)
\Align[l] (or we return on $Nop()$ case.) (35,-10)
\cip
$$
\begin{itemize}
\item $op_1'=Nop()$ : 
The number of nodes are different, therefore $t_1'\neq t_2'$
\item $op_1'=Add(\_,\_)$ idem
\item $op_1'=Del(\_,\_)$ idem
\end{itemize}
%
\end{itemize}
$\Box$
\end{proof}

\subsection{A Stronger Deletion ensures TP1, TP2 for trees}
\label{subsec:delete2-is-TP1-TP2}
\subsubsection{The New Set of Operations and $\IT$.}


Let us define a new deletion operation.

$Del_2(p,n)$ : Delete the subtree accessed from the edge  labeled by $n$  at the
end of path $p$. 
$$\begin{array}{l}
Del_2(n'.p,n)(t)=t, $ if $ n\not \in Dom(t)\\
Del_2(n_i.p,n)(\{n_1(t_1),...,n_i(t_i),...,n_q(t_q)\}) =\\
\hspace{4cm} \lmult n_1(t_1),...,n_i(Del_2(p,n)(t_i)),...,n_q(t_q)\rmult \\
Del_2(\epsilon,n)(t)=t , $ if $ n\not \in Dom(t)\\
Del_2(\epsilon,n_i)(\{n_1(t_1),...,n_i(t_i),...,n_q(t_q)\})=
         \lmult n_1(t_1),...,n_q(t_q)\rmult
\end{array}$$

Let $\Op$ be the set of operations
$\{Nop(),Add(p,n),Del_2(p,n)\}$ $p\in
\mathcal{P}, n \in \Sigma$ and let $IT$ be defined by:

The $\IT$ function is defined by:
\[
IT(op_1,op_2)= \\ \\\left \{ \begin{array}{l}
IT(Add(p,n),Add(p',n'))=
Add(p,n), \\
IT(Add(p,n),Del(p',n'))=\left \{ \begin{array}{l} 
Nop() ,$ if $p=p' \wedge n=n'\\
Nop(), $ if $ p'.n' \triangleleft p \\
Add(p,n) , $ else. $\\
\end{array} \right.\\ 
IT(Del(p,n),Add(p',n'))=Del(p,n)\\
IT(Del(p,n),Del(p',n'))=\left \{ \begin{array}{l} 
Nop(), $ if $ p=p' \wedge n=n'\\
Nop(), $ if $ p'.n' \triangleleft p\\
Del(p,n), $ else.  $
\end{array} \right.\\
IT(op_1,Nop())=op_1\\
IT(Nop(),op_2)=Nop();\\
\end{array} \right.
\]

\subsubsection{Proof of TP1 and TP2 with Strong Deletion} 

\begin{theorem}
\label{thm:trees-IT-TP1-TP2}
$IT$ satisfies $TP1$ and $TP2$.
\end{theorem}

\paragraph{Proving TP1}
\label{TP1}
$\forall op_1,op_2 \in Op, s \in State ,$\\$ (t)[op_1;IT(op_2,op_1)]=(t)[op_2;IT(op_1,op_2)]$\\

We perform a case analysis on $op_1$ and $op_2$ : 

\begin{enumerate}
\item $op_1=Add(p,n)$ and $op_2=Add(p',n')$\\
$(t)[Add(p,n);IT(Add(p',n'),Add(p,n))] = $\\ $ (t)[Add(p,n);Add(p',n')]$\\\\
$(t)[Add(p',n');IT(Add(p,n),Add(p',n'))] = $ \\$ (t)[Add(p',n');Add(p,n)]$

We prove : \\
$Do(Do(t,Add(p,n)),Add(p',n')) = $\\$Do(Do(t,Add(p',n')),Add(p,n))$.\\

We  perform an induction on path length.

\begin{enumerate}

\item Empty path : 
\begin{itemize}

\item If $n,n' \not \in Dom(t)$ and $n\neq n'$\\
$Add(\epsilon,n')(Add(\epsilon,n)(\{n_1(T_1),...,n_q(T_q)\}))$\\
$=\{n_1(T_1),...,n_q(T_q),n(\{\}),n'(\{\})\}$\\
$Add(\epsilon,n')(Add(\epsilon,n)(\{n_1(T_1),...,n_q(T_q)\}))$\\
$=\{n_1(T_1),...,n_q(T_q),n'(\{\}),n(\{\})\}$\\
which  are equal.

\item If $n=n'$ \\
We obtain : $=\{n_1(T_1),...,n_q(T_q),n(\{\})\}$\\
Because we use the third choice of function $Add(\epsilon,n)(t)$ and first operation add $n(\{\})$.

\item If $n\in Dom(t)$ \\
We have $\{n_1(T_1),...,n_q(T_q),n'(\{\})\}$
Third we use the second case of definition

\item idem if $n'\in Dom(t)$ with $n$.

\item If $n,n'\in Dom(t)$ the tree is unchanged.
\end{itemize}

\item if $p.n \triangleleft p' : \exists p'', p.n.p''=p$
if $n \in dom(p)$ then $Add(p,n)$ do nothing.\\
else \\
We have\\
 $\proj{p}{t}=\{n_1(T_1), ... , m_1(T'_1), ... n_{q}(T_{q})\}$
$^{(1)} = \{n_1(T_1), ... , m_1(T'_1), ... n_{q}(T_{q}), n(\{\})\}\\$
$\{n_1(T_1), ... , m_1(T'_1), ... n_{q}(T_{q}), n(Add(p'',n')(\{\})\}$\\

By definition : \\
$Add(p',n')(t)=\{n_1(T_1), ... , m_1(T'_1), ...,\\ n_{q}(T_{q}, n(Add(p'',n')(\{\})\}$
therefor $n \in dom(\proj{p}{t})$ and $Add(p,n)$ do nothing.
\\so $^{(2)}=\{n_1(T_1), ... , m_1(T'_1), ... n_{q}(T_{q}), n(Add(p'',n')(\{\})\}$

\item idem for  $p'.n' \triangleleft p $

\item if $p \triangleleft p'$
We have : $p'=p.m_1.p_1$\\
We have $\proj{p}{t}=\{n_1(T_1), ... , m_1(T'_1), ..., n_{q-1}(T_{q-1})\}$\\

Two cases occurs, by recurrence definition : \\$\proj{p}{t'}=\\\{n_1(T_1), ... , m_1(Add(p'_1,n')(T'_1)), ..., n_{q-1}(T_{q-1}), n(\{\})\}$
\item idem for $p' \triangleleft p$

\item {$p,p'$ not empty (general case)}
\label{generalcase}
$\exists p \in \mathcal{P} | p=p_{comon}.p_1'$ and $p'=p_{comon}.p_2'$
We have two non-empty paths then : \\
$p_1'=m_1.p_1''$ and $p_2'=m_2.p_2''$\\

We have\\
$\proj{p}{t}=\{n_1(T_1), ... , m_1(T'_1), ... , m_2(T'_2), ... n_{q-2}(T_{q-2})\}$\\
We have by definition : \\$\proj{p}{t'}=\{n_1(T_1), ... , m_1(Add(p_1'',n)(T'_1), ... $\\$, m_2, (Add(p''_2,n')(T_2)), ..., n_{q-2}(T_{q-2})\}$
\end{enumerate}
\item $op_1=Add(p,n)$ and $op_2=Del(p',n')$\\
$(t)[Add(p,n);IT(Del(p',n'),Add(p,n))] ^{(1)} $\\
$(t)[Del(p',n');IT(Add(p,n),Del(p',n'))] ^{(2)}$
\begin{itemize}
\item $p=p'$ and $n=n'$\\
$^{(1)}=(t)[Add(p,n),Del(p,n)]$\\
$^{(2)}=(t)[Del(p,n; Nop()]$
\begin{itemize}
\item if $n \in Dom(p) $ then $Add(p,n)(t)$ do nothing.
Therefore $^{(1)}=^{(2)}$
\item if $n\not \in Dom(p) $ then $Add(p,n)(t)$  create a node who delete by $Del(n,p)$ in $^{(1)}$.
and $Del(n,p)$ do nothing in $^{(2)}$\\
Therefore $^{(1)}=^{(2)}$
\end{itemize}
\item $p'.n'\triangleleft p$\\
$^{(1)} =  (t)[Add(p,n);Del(p',n')]$\\
$^{(2)} =  (t)[Del(p',n');Nop()]$\\
if p=p'.n'.p''\\
We take : $\proj{p'}{t} = \{n_1(T_1), ...,n'(T),..., n_{q-1}(T_{q-1})\}$
We have :\\$\proj{p}{^{(1)}}= \{n_1(T_1), ...,n'(Add(p'',n)(T)),..., n_{q-2}(T_{q-2})\}$
$= \{n_1(T_1), ..., n_{q-2}(T_{q-2})\}$\\
$\proj{p}{^{(2)}}= \{n_1(T_1), ..., n_{q-2}(T_{q-2})\}$
\item else : same demo of \ref{generalcase}.
\end{itemize}
\item idem for $op_1=Del(p,n)$ and $op_2=Add(p',n')$
\item $op_1=Del(p,n)$ and $op_2=Del(p',n')$\\
$(t)[Del(p,n);IT(Del(p',n'),Del(p,n))] ^{(1)} $\\
$(t)[Del(p',n');IT(Del(p,n),Del(p',n'))] ^{(2)}$
\begin{itemize}
\item p.n=p'.n' :\\
$^{(1)}=(t)[Del(p,n),Nop()]$\\
$^{(2)}=(t)[Del(p,n),Nop()]$
\item $p.n \triangleleft p'$\\
We have p'=p.n.p'';\\
We take $\proj{p}{t}=\{n_1(T_1),...,n(T),...,n_{q-1}(T_{q-1})\}$\\
$^{(1)}=(t)[Del(p,n);Nop()] $\\
$^{(2)}=(t)[Del(p',n');Del(p,n)] $\\
$\proj{p}{^{(1)}}=\{n_1(T_1),...,n_{q-1}(T_{q-1})\}$

first time : $\proj{p}{Del(p,n)(t)}\{n_1(T_1),...,\spliteqn n(Del(p'',n')(T)),...,n_{q-1}(T_{q-1})\}$\\
therefore $\proj{p}{^{(2)}}=\{n_1(T_1),...,n_{q-1}(T_{q-1})\}$
\item idem for $p'.n' \triangleleft p'$
\item else: same \ref{generalcase} we have two independant subtree.
\end{itemize}

\item case Nop() is trivial. \hfill $\Box$
\end{enumerate}

\paragraph{Proving TP2}
\label{TP2}
$IT(IT(Op,Op_1),IT(Op_2,Op_1))^{(1)} =$\\$ IT(IT(Op,Op_2),IT(Op_1,Op_2))^{(2)}$\\
We will explore every case :
\begin{itemize}
\item $Op=Add(p,n)$, $Op_1=Add(p_1,n_1)$  and $Op_2=Add(p_2,n_2)$
therefore $^{(1)}=Add(p,n)$ and $ ^{(2)}=Add(p,n)$
\item $Op=Add(p,n), Op_1=Add(p_1,n_1)$  and $Op_2=Del(p_2,n_2)$\\
$IT(IT(Add(p,n),Add(p_1,n_1)),IT(Del(p_2,n_2),Add(p_1,n_1)))^{(1)}$\\
    $IT^{(b)}(IT(Add(p,n),Del(p_2,n_2), \spliteqn IT^{(a)}(Add(p_1,n_1),Del(p_2,n_2)))^{(2)}$ \\
$^{(1)}=IT(Add(p,n),Del(p_2,n_2))$\\
$^{(2)}=IT^{(b)}(IT(Add(p,n),Del(p_2,n_2)),Add(X,n_1))=IT(Add(p,n),Del(p_2,n_2))$\\
or $^{(2)}=IT^{(b)}(IT(Add(p,n),Del(p_2,n_2)),Nop())=$\\$IT(Add(p,n),Del(p_2,n_2))$\\
Because $^{(a)}$ give a Add() or a Nop() the second argument of $^{(b)}$ is a $Add$ or a $Nop$.
%
\item Idem for  $Op=Add(p,n), Op_1=Del(p_1,n_1)$  and $Op_2=Add(p_2,n_2)$
\item  $Op=Add(p,n), Op_1=Del(p_1,n_1)$  and $Op_2=Del(p_2,n_2)$\\
$IT(IT(Add(p,n),Del(p_1,n_1)),IT(Del(p_2,n_2),Del(p_1,n_1)))^{(1)}$\\
$IT(IT(Add(p,n),Del(p_2,n_2)),IT(Del(p_1,n_1),Del(p_2,n_2)))^{(2)}$
\begin{itemize}
\item If $p_1=p_2$ and $n_1=n_2$ \\
$^{(1)}=IT(IT(Add(p,n),Del(p_1,n_1)),Nop())$\\
$^{(2)}=IT(IT(Add(p,n),Del(p_1,n_1)),Nop())$
\item If $p_2.n_2 \triangleleft p_1$\\
$^{(1)}=IT(IT(Add(p,n),Del(p_1,n_1)),Del(p_2,n_2)))$\\ because $p_1.n_1 \neq p_2.n_2$\\
$^{(2)}=IT(IT(Add(p,n),Del(p_2,n_2)),Nop())$
\begin{itemize}
\item if $p_2.n_2 \triangleleft p$\\
$^{(1)}=IT(IT(Add(p,n),Del(p_1,n_1)),Del(p_2,n_2)))$\\
$^{(2)}=Nop()$
\begin{itemize}
\item if $p_1.n_1 \triangleleft p$\\
$^{(1)}=Nop()$\\
$^{(2)}=Nop()$
\item else : \\
$^{(1)}=IT(Add(p,n),Del(p_2,n_2))=Nop()$\\
$^{(2)}=Nop()$
\end{itemize}
%
\item idem for $p_1.n_1 \triangleleft p$
\item else : \\
$^{(1)}=IT(Add(p,n),Del(p_2,n_2))=Add(p,n)$ because\\ $p_2.n_2 \not\!\triangleleft p \wedge p_1.n_1 \not\!\triangleleft p$ \\
$^{(2)}=IT(IT(Add(p,n),Del(p_2,n_2)),Nop()))=^{(1)}$
\end{itemize}
\item idem if $p_1.n_1\triangleleft p_2$
\item Else :\\
$^{(1)}=IT(IT(Add(p,n),Del(p_1,n_1)),Del(p_2,n_2)))$\\
$^{(2)}=IT(IT(Add(p,n),Del(p_2,n_2)),Del(p_1,n_1)))$\\
\begin{itemize}
\item if $p=p_1 \wedge n=n_1$\\
$^{(1)}=IT(Nop(),Del(p_2,n_2)))=Nop()$\\
$^{(2)}=IT(IT(Add(p_1,n_1),Del(p_2,n_2)),Del(p_1,n_1)))$\\
By hypothese : \\ $^{(2)}=IT(Add(p_1,n_1),Del(p_1,n_1)))=Nop()$
\item idem if $p=p_2 \wedge n=n_2$
\item if $p_1.n_1 \triangleleft p$
$^{(1)}=IT(Nop(),Del(p_2,n_2)))$\\
$^{(2)}=IT(IT(Add(p,n),Del(p_2,n_2)),Del(p_1,n_1)))$\\

We have
 $p_2\neq p \vee n_2 \neq n$ and $p_2.n_2 \not \triangleleft p$
 because 

$p_2.n_2 \triangleleft p \wedge p_1.n_1 \triangleleft p \Rightarrow
p_1.n_1\triangleleft p_2.n_2 \vee p_2.n_2\triangleleft p_1.n_1$\\
$^{(2)}=IT(Add(p,n),Del(p_1,n_1)))=Nop()=^{(1)}$
\item idem if $p_2.n_2 \triangleleft p$
\item else : \\
$^{(1)}=IT(Add(p,n))$\\
$^{(2)}=IT(Add(p,n))$
\end{itemize}
\end{itemize}
\item  Trivial for $Op=Del(p,n), Op_1=Add(p_1,n_1)$ and $Op_2=Add(p_2,n_2)$
\item if $Op=Del(p,n), Op_1=Del(p_1,n_1)$ and $Op_2=Add(p_2,n_2)$\\

$^{(1)}=IT(IT(Del(p,n),Del(p_1,n_1)),\spliteqn IT(Add(p_2,n_2),Del(p_1,n_1))$\\
$^{(1)}=IT(Del(p,n),Del(p_1,n_1))$ because the first argument will be a
'$Del$' 
and the second will be a '$Add$'. \\
$^{(2)}=IT(IT(Del(p,n),Add(p_2,n_2)),\spliteqn IT(Del(p_1,n_1),Add(p_2,n_2))$\\
$^{(2)}=IT(Del(p,n),Dell(p_1,n_1))$

\item idem for $Op=Del(p,n), Op_1=Add(p_1,n_1)$ and $Op_2=Del(p_2,n_2)$

\item if $Op=Del(p,n), Op_1=Del(p_1,n_1)$ and $Op_2=Del(p_2,n_2)$

\item Trivial If $Op=Nop()$

\item if $Op=X(n,p), Op_1=Nop()$ and $Op_2=X'(n_2,p_2)$\\
$^{(1)}=IT(IT(X(p,n),Nop()),IT(X'(p_2,n_2),Nop())$\\
$=IT(X(p,n),X'(p_2,n_2))$\\
$^{(2)}=IT(IT(X(p,n),X'(p_2,n_2)),IT(Nop(),X'(p_2,n_2))$\\
$=IT(X(p,n),X'(p_2,n_2))$

\item idem $Op=X(n,p), Op_1=X(p_1,n_1)$ and $Op_2=Nop()$

\item Trivial, if $Op=X(n,p) , Op_1=Nop()$ and $Op_2=Nop()$
\end{itemize}
\hfill $\Box$

\subsection{Proof of Theorem \ref{thm:IT-is-TP1-TP2}}

The proof is similar to the previous proof and has been checked by Vote using
the following specification:

\begin{verbatim}
%VOTE file for proving TP1/TP2 on XML like trees

type node(mem,data),lbl(novalue),nat;
observator
%test node existence
 bool exist(node);
%relation between son and father
 bool childof(node, node);
%returns the label of a node
 lbl getLbl(node);
auxiliary 
%returns tree if there is a path between nodes
 bool childofp(node, node);

operation
%add a node n, if it doesn't exists, 
%it becomes a son of p that must exist
 not(exist(n)) and exist(p) and (n!=mem) and (n!=data)		: Add(node p,node n);

%delete an existing node that must be different 
%from the two initial nodes mem and data
 exist(n) and (n!=mem) and (n!=data)				: Del(node n);

%site t moves node n under node p if n exists and is different 
%from mem and data
exist(n) and exist(p) and (n!=mem) and (n!=data) and (n != p): Move(node n, node p,nat t);

%site t renames a node  n with  label l if n exists and is 
%different from mem and data
 exist(n) and (n!=mem) and (n!=data): Ren(node n,lbl l,nat t);

transform
%definition of the IT transformation
 T(Add(p1,n1),Del(n2)) = if (p1==n2) then 
		return Add(mem,n1)
	else 
		return Add(p1, n1)
	endif;
T(Ren(n1,l1,s1), Del(n2)) = if (n1==n2) then 
		return nop
	else
		return Ren(n1,l1,s1)
	endif;

T(Ren(n1,l1,s1),Ren(n2,l2,s2))= if (n1==n2 and s1 > s2) then
		return nop
	else
		return Ren(n1,l1,s1)
	endif;

T(Move(n1,p1,s1),Move(n2,p2,s2)) = if (n1==n2 and s1 > s2) then
		return nop
	else
		return Move(n1,p1,s1)
	endif;
T(Move(n1,p1,s1),Del(n2)) = if(n1==n2) then 
		return nop
	elseif (p1 == n2) then
		return Move(n1,mem,s1)
	else 
		return Move(n1,p1,s1)
	endif;
definition
exist'(n1)/Add(p2,n2) = if (n1 == n2) then return true
				elseif (n1==mem or n1==data) then
					return true
				else return exist(n1)
				endif;
exist'(n1)/Del(n2) =  if(n1==mem or n1==data) then
					return true
			elseif (n1 == n2) then return false
 				else return exist(n1)
				endif;
childof'(n1,p1)/Add(p2,n2) = if (n1 == n2 and p2==p1) then return true
					else return childof(n1,p1)
					endif;
childof'(n1,p1)/Del(n2) = if (n2 == n1) then 
					return false
				elseif (n2==p1) then 
					return false
				elseif (p1==mem and childof(n1,n2)) then
					return true
 				else 
					return childof(n1,p1)
				endif;
				
childof'(n1,p1)/Move(n2,p2,s1) = if(n1 == n2 and p1==p2) then 
					return true
				elseif (n1==n2 and p1!=p2) then 
					return false
				else 
					return childof(n1,p1)
				endif;
getLbl'(n1)/Add(p2,n2) = if (n1==n2) then return novalue 
 				else
					return getLbl(n1)
 				endif;
getLbl'(n1)/Del(n2) = if (n2 == n1) then return novalue
 				else
					return getLbl(n1)
				endif;
getLbl'(n1)/Ren(n2,l2,s2) =if (n1==n2) then
					return l2
 				else
					return getLbl(n1)
				endif;
lemma
%basic lemmas needed for the proof

%all trees have node meme and data
=>exist(mem);
=>exist(data);

%assume no auto-concurrency
s1>=s2 and s2>=s1 =>;
not( s1>s2) and not(s2>s1) =>;

%Axioms for trees
childof(x,y) and childof(x,z) and (z!=y) =>;
childofp(x,y) and childofp(y,z)=>childofp(x,z);
childof(x,y)=> childofp(x,y);
childofp(x,x)=>;
\end{verbatim}

The output of Vote is:

\begin{verbatim}
Elapsed time: -704.857296 s

--- Global statistics of the main successful operations ---

- contextual_rewriting   : 0 of 0 tries.
- equational_rewriting   : 0 of 0 tries.
- conditional_rewriting  : 334 of 85455 tries.
- partial_case_rewriting : 0 of 0 tries.
- total_case_rewriting   : 675 of 675 tries.
- induction              : 0 of 0 tries.
- subsumption            : 165 of 63888 tries.
- tautology              : 71 of 245580 tries.

-----------
  Total clauses: 30428

  Max depth    : 1

All sets of conjectures were successfully processed
\end{verbatim}

\subsection{Proof of Theorem \ref{thm:combination}}

We give the proof of the combination theorem.
\begin{proof}
Given a sequence of computations $Comp$ i.e. a sequence of expressions
$\phi_l(E).!r_l$ or $r_e.\phi_e(E)$ respecting causality, we extract $Comp_T$
and $Comp_{id}$ the respective computations of $\A_T$ and $A_D$ for each $id$:

$
\begin{array}{l}
\Pi_T(\phi_l(\pair{E_T,E_D}).!r_l)= \varphi^T_l(E_T).!r_l^T$ if $r_l=\pair{r_l^T,\bot}\\
\Pi_{id}(\phi_l(\pair{E_T,E_D}).!r_l)=\varphi^D_l(E_D(id)).!r_l^D$ if $r_l=\pair{\bot,(id,r_l^D)}\\
\Pi_T(?r_e.\phi_e(\pair{E_T,E_D})=?r_e^T.\varphi^T_l(E_T)$ if $r_e=\pair{r_e^T,\bot}\\
\Pi_{id}(?r_e.\phi_e(\pair{E_T,E_D})=?r_e^D.\varphi_e^D(E_D(id))$ if $r_e=\pair{\bot,(id,r_e^D)}
\end{array} 
$\\
and for all other cases $\Pi_T(\ldots)=\Pi_{id}(\ldots)=0$ where $0$ is the null
process that does nothing. By construction $Comp_T$ respects the causality
relations restricted to the operations of $\Op_T$. The same holds for $Comp_{id}$.
the causality relation (but the reverse doesn't necessarily holds). Therefore
$Comp_T$ is a legal computation of $\A_D$ and by the convergence of $\A_D$
each site has he same state $s_T$. For each $id\in Dom(s_T)$, the sequence
$Comp_{id}$ is  legal computation of $\A_{id}$, therefore each site has the
same state $s_{id}$. 
\end{proof}

\end{document}

%% file: macros.tex
\newcommand\keywords[1]{{\bf Keywords:} #1}
\newcommand\text[1]{#1}

\def\ra{\rightarrow}

\def\Name{\mathcal{N}}

\def\Op{Op}
\def\OP{Op}
\def\ID{ID}
\def\T2XML{\varphi}
\def\ITS{\IT^{*}}

\def\TP2S{TP2^{*}}

\newcommand{\proj}[2]{{#2}_{\mid_{#1}}}
\newcommand{\projb}[2]{{#2}_{\lceil_{#1}}}
\newcommand{\lmult}{\{}
\newcommand{\rmult}{\}}
\newcommand{\mult}[1]{\lmult #1 \rmult}
\newcommand{\data}{data}
\newcommand{\mem}{mem}
\newcommand{\idp}{id_p}
\def\spliteqn{\\  }
\def\munion{\oplus}        

\def\et{\wedge}
\def\non{\neg}

\def\IT{IT}

\def\Pos{Pos}

\newcommand\pair[1]{\langle #1 \rangle}
\newcommand\req[1]{\langle #1 \rangle}
\def\ENVT{\mathcal{E}nv_T}
\def\REQT{\mathcal{R}eq_T}
\def\ENVD{\mathcal{E}nv_D}
\def\REQD{\mathcal{R}eq_D}
\def\ENV{\mathcal{E}nv}
\def\REQ{\mathcal{R}eq}

\def\A{\mathcal{A}}

%% file: tp1.pstex_t
\begin{picture}(0,0)%
\includegraphics{tp1.pstex}%
\end{picture}%
\setlength{\unitlength}{3108sp}%
\begingroup\makeatletter\ifx\SetFigFontNFSS\undefined%
\gdef\SetFigFontNFSS#1#2#3#4#5{%
  \reset@font\fontsize{#1}{#2pt}%
  \fontfamily{#3}\fontseries{#4}\fontshape{#5}%
  \selectfont}%
\fi\endgroup%
\begin{picture}(3540,981)(121,-454)
\put(136,-106){\makebox(0,0)[lb]{\smash{{\SetFigFontNFSS{9}{10.8}{\rmdefault}{\mddefault}{\updefault}{\color[rgb]{0,0,0}$t$}%
}}}}
\put(586,254){\makebox(0,0)[lb]{\smash{{\SetFigFontNFSS{9}{10.8}{\rmdefault}{\mddefault}{\updefault}{\color[rgb]{0,0,0}$op_1$}%
}}}}
\put(586,-376){\makebox(0,0)[lb]{\smash{{\SetFigFontNFSS{9}{10.8}{\rmdefault}{\mddefault}{\updefault}{\color[rgb]{0,0,0}$op_2$}%
}}}}
\put(2746,209){\makebox(0,0)[lb]{\smash{{\SetFigFontNFSS{9}{10.8}{\rmdefault}{\mddefault}{\updefault}{\color[rgb]{0,0,0}$t'_1$}%
}}}}
\put(2746,-286){\makebox(0,0)[lb]{\smash{{\SetFigFontNFSS{9}{10.8}{\rmdefault}{\mddefault}{\updefault}{\color[rgb]{0,0,0}$t'_2$}%
}}}}
\put(3646,-106){\makebox(0,0)[lb]{\smash{{\SetFigFontNFSS{9}{10.8}{\rmdefault}{\mddefault}{\updefault}{\color[rgb]{0,0,0}$TP1$ implies $t'_1=t'_2$}%
}}}}
\put(1576,-106){\makebox(0,0)[lb]{\smash{{\SetFigFontNFSS{9}{10.8}{\rmdefault}{\mddefault}{\updefault}{\color[rgb]{0,0,0}$IT(op_1,op_2)$}%
}}}}
\put(1531,344){\makebox(0,0)[lb]{\smash{{\SetFigFontNFSS{9}{10.8}{\rmdefault}{\mddefault}{\updefault}{\color[rgb]{0,0,0}$IT(op_2,op_1)$}%
}}}}
\put(1216,164){\makebox(0,0)[lb]{\smash{{\SetFigFontNFSS{9}{10.8}{\rmdefault}{\mddefault}{\updefault}{\color[rgb]{0,0,0}$t_1$}%
}}}}
\put(1216,-286){\makebox(0,0)[lb]{\smash{{\SetFigFontNFSS{9}{10.8}{\rmdefault}{\mddefault}{\updefault}{\color[rgb]{0,0,0}$t_2$}%
}}}}
\end{picture}%